\documentclass[12pt]{article}
\usepackage{amsmath,amssymb,epsfig}
\usepackage{color}
\input{colordvi.tex}
\def\unit{{\relax{\rm 1\kern-.26em I}}}

\newcommand{\tr}{{\rm Tr}}

\makeatletter

\renewcommand\section{\@startsection {section}{1}{\z@}%
                                   {-3.5ex \@plus -1ex \@minus -.2ex}%
                                   {2.3ex \@plus.2ex}%
                                   {\normalfont\large\bfseries}}

\renewcommand\subsection{\@startsection{subsection}{2}{\z@}%
                                     {-3.25ex\@plus -1ex \@minus -.2ex}%
                                     {1.5ex \@plus .2ex}%
                                     {\normalfont\normalsize\bfseries}}

\makeatother

\begin{document}

% format
\baselineskip=18pt  % a la harvmac
\numberwithin{equation}{section}  % make eq labels (sec.num)
\allowdisplaybreaks  % allow page breaks in displayed eqs

% print date, time and filename
%\pagestyle{myheadings}
%\markright{{\tt \jobname.tex} -- \today{} \now}

%%%%%%%%%%%%%%%%%%%%%%%%%%%%%%%%%%%%%%%%%%%
%%%        TITLE BEGINS HERE
%%%%%%%%%%%%%%%%%%%%%%%%%%%%%%%%%%%%%%%%%%%

%% ========== title (note version) begins here ==========
%
%\vspace*{-1cm}
%\begin{center}
% {\Large\bf Title of the Document}
%\end{center}
%\vspace*{-.5cm}
%
%% ========== title (note version) ends here ==========

%% ========== title (paper version, a la harvmac) begins here ==========

\thispagestyle{empty}

% Report number
\vspace*{-2cm}
\begin{flushright}
%{\tt arXiv:yymm.nnnn}\\
\end{flushright}

\begin{flushright}
%MCTP-XX-XX\\
KYUSHU-HET-153
\end{flushright}

\begin{center}

\vspace{1.4cm}

%\vspace{0.5cm}
%{\bf\Large Fuzzy Monopole, Fuzzy String and }

\vspace{1cm}
{\bf\Large Gravitational Correction to Fuzzy String}
\vspace*{0.3cm}

{\bf\Large  in Metastable Brane Configuration}
\vspace*{0.2cm}

\vspace{1.3cm}

{\bf
Aya Kasai$^{1}$ and  Yutaka Ookouchi$^{2,1}$} \\
\vspace*{0.5cm}

${ }^{1}${\it Department of Physics, Kyushu University, Fukuoka 810-8581, Japan  }\\
${ }^{2}${\it Faculty of Arts and Science, Kyushu University, Fukuoka 819-0395, Japan  }\\

\vspace*{0.5cm}

\end{center}

\vspace{1cm} \centerline{\bf Abstract} \vspace*{0.5cm}

We study dynamics of a cosmic string in a metastable brane configuration in Type IIA string theory. We first discuss a decay process of the cosmic string via a fuzzy brane (equivalently bubble/string bound state) by neglecting gravitational corrections in ten-dimension. We find that depending on the strength of the magnetic field induced on the bubble, the decay rate can be either larger or smaller than that of $O(4)$ symmetric bubble. Then, we investigate gravitational corrections to the fuzzy brane by using the extremal black $NS$-five brane solution, which makes the lifetime of the metastable state longer.

\newpage
\setcounter{page}{1} % don't number title page

%% ========== title (paper version, a la harvmac) ends here ==========

%%%%%%%%%%%%%%%%%%%%%%%%%%%%%%%%%%%%%%%%%%%
%%%           TITLE ENDS HERE
%%%%%%%%%%%%%%%%%%%%%%%%%%%%%%%%%%%%%%%

%%%%%%%%%%%%%%%%%%%%%%
\section{Introduction}

Remarkable progress on string compactification may suggest us that the
potential of string theories has a complicated structure and admits a large number of metastable vacua \cite{Lands}. This
string landscape, if it is true, would open up a new avenue for
cosmology at the early stage of the universe. One of interesting dynamics in the
string landscape is transition between various metastable vacua (see \cite{Bubble1,Bubble2,Bubble3,Bubble4} and references therein for earlier works). It is
fascinating to think that transitions between vacua occur
several times at the early stage and eventually the system arrives at
the present universe with small cosmological constant. In this sense,
studies of the lifetime of various vacua would be important subject. However due to lack of full knowledge of the string landscape, analytic studies on this subject are quite hard. So it would be useful to take a limit in which we can neglect gravitational corrections and to extract lessons on longevity of vacua from ``limited'' string landscape.  

In the first paper \cite{KO}, the authors studied inhomogenious vacuum
decay via a stringy monopole. The authors showed that dielectric
branes \cite{Myers}, which are bound states of a spherical $D5$ brane and a $D3$
brane, are key objects for decays of false vacua. It can be
either unstable or metastable depending on the magnetic flux
originating from the dissolving $D3$ brane in the $D5$ brane. As
emphasized, the lower energy vacuum filling in the bubble offers a
force to enlarge the radius of the spherical bubble and in fact,
a stable fuzzy monopole with finite size (dielectric brane) was formed without using background flux  \cite{Myers,Emparan} nor angular momentum \cite{Supertube}. This is a new mechanism for creating stable fuzzy branes. Also, when the induced magnetic flux is large enough, the fuzzy monopole becomes unstable and expands its radius without bound. This is remarkable because the lifetime of the false vacuum becomes drastically shorter due to the decay of the unstable monopole. A decay process exploiting a fuzzy brane was initially studied in \cite{KPV} where anti-$D3$ branes create a fuzzy $NS5$ brane, and later extended to various decay channels in related models \cite{Frey}. Recent progress on this subject can be seen in \cite{newKPV} and references therein.

The idea of the inhomogeneous decay of false vacua via solitons was initially pointed out by \cite{First1,First2,First3}. Then, in \cite{KumarYajnik,OurRstring,LeeYajnik}, applications to phenomenological model building were discussed in the context of field theories. Our first paper \cite{KO} can be regarded as the first realization of this idea in string theories. In this paper, we would like to go a step further toward including gravitational effects. To this goal, we use a brane configuration in Type IIA string theory. In this context, it is relatively easy to incorporate the gravitational corrections by means of replacing an $NS$-five brane with the extremal black-brane solution, which allows us to evaluate the decay rate explicitly within the validity of the brane-limit. This is one of advantages to use a metastable brane configuration to get an insight into decay processes under gravitational effects.

The plan of this paper is as follows. In section 2, we review a metastable configuration in Type IIA string theory \cite{OoguriOokouchi,Franco,IAS,Giveon}. In section 3, after reviewing a cosmic string existing in the metastable configuration along the lines of \cite{EHT}, we investigate stability of a dielectric brane which is a bound state of the string and a tube-like domain wall connecting the false and true vacua. In section 4, we discuss gravitational corrections to the decay process affected by the $NS$-five brane background. Section 5 is devoted to conclusions and discussions.

%%%%%%%%%%%%%%%%%%%
\section{Set-up of brane configuration}

In this section, we quickly review false and true vacua in brane configuration in Type IIA string theory discussed in
\cite{OoguriOokouchi,Franco,IAS,Giveon}. See
\cite{KOOreview} for a
recent review on
supersymmetry (SUSY)
breaking vacua in string
theories. In contrast to the
work \cite{KO}, the potential barrier between
false and true vacua
is very shallow, because it is created by higher order
corrections. Hence, estimation of the decay rate of
the false vacuum is carried out by
triangle approximation
\cite{triangle} rather than
thin-wall approximation \cite{Coleman},
which leads to a slightly
different conclusion from
\cite{KO}: existence of
a soliton does not necessarily
make the lifetime of
false vacuum shorter.

To begin with, let us review the supersymmetry breaking brane configuration \cite{OoguriOokouchi,Franco,IAS,Giveon}. It is useful to introduce the following parametrization for internal space,
\begin{equation}
v=x^4+ix^5,\qquad w=x^8+ix^9,\qquad y=x^6.
\end{equation}
Consider three $NS$-five branes extending in $v$ or $w$ directions. We refer to branes placed at $(y,v)=(y_1,0)$, $(y,v)=(y_1+\Delta y,\Delta v)$ and $(y,w)=(y_{NS},0)$ as $NS_1$, $NS_2$ and $NS_3$ branes respectively. There are $N$ $D4$ branes stretched between the $NS_2$ and $NS_3$ branes. Also, $n$ tilted $D4$ branes are stretched between the $NS_1$ and $NS_2$ branes.

%%%%%%%%%%%%%%%%%%%%%%%%%%%%
\begin{figure}[htbp]
\begin{center}
 \includegraphics[width=.5\linewidth]{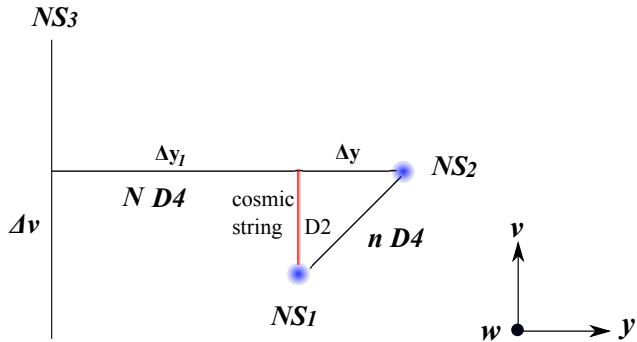}
\vspace{-.1cm}
\caption{\sl A metastable brane configuration. As we will discuss in the next section, a $D2$ brane filling $x^4$ direction in $v$ plane corresponds to a cosmic string \cite{EHT}. We define $\Delta y_1=y_1-y_{NS}$ for later convenience.}
\label{brane1}
\end{center}
\end{figure}
%%%%%%%%%%%%%%%%%%%%%%%%%%%%

If we take the field theory limit, $l_s\to 0$, while keeping the following quantities finite,
\begin{equation}
 h^2={8\pi^2g_sl_s\over \Delta y } ,\qquad \mu^2={\Delta v\over 16\pi^3 g_s l_s^3},
\end{equation}
this brane configuration can be interpreted as the $U(N)$ gauge theory with $N+n$ flavors and the following superpotential \cite{ISS},
\begin{equation}
W_{\rm  mag}=h  q^j_\alpha \Phi^i_j \tilde q_i^\alpha- h \mu^2 \Phi_i^i,\label{magPotential}
\end{equation}
where $\Phi$ is a singlet in the $U(N)$ gauge group. $i,j$ run from
$1$ to $N+n$ and $\alpha$ is the color index. The vacuum energy in
this configuration is\footnote{Note that we included modifications of vacuum energy coming from higher order corrections in the definition of the coupling constant $h$. }
\begin{equation}
V_{\rm meta}=n |h\mu^2|^2.
\end{equation}
Note that in this brane configuration, there is an $n\times n$ matrix
$X$ of massless fields corresponding to the positions of the tilted
$D4$ branes in the $w$ direction. This is a flat direction of this
brane configuration at tree level. Hereafter we assume $n=1$ for the
sake of simplicity. As discussed in \cite{ISS,Kutasov}, there are two
kinds of corrections which lift the flat direction. One is the
Coleman-Weinberg (CW) potential generated by open strings connecting
between the $N$ $D4$ and the $n$ tilted $D4$ branes. The other comes from the
gravitational effect originating from $NS_3$ brane as we will discuss
in section 4 in detail. The second contribution becomes relevant in
the so-called brane limit where $g_s$ is very small but $l_s$ is
finite. In \cite{ISS,Kutasov}, the explicit calculation has been done,
\begin{eqnarray}
V_{\rm correction}&=&\left(  {\ln 4-1 \over 8\pi^2}N |h^2\mu|^2   +{\Delta v \over \sqrt{(\Delta y)^2+(\Delta v)^2}}{l_s\over y_{NS}^2}  \right)  \tr X^{\dagger}X+\cdots  \nonumber \\
&=& K_{0} |h^2\mu|^2 \tr X^{\dagger}X+\cdots ,
\end{eqnarray}
where we have defined  
\begin{equation}
K_{0}\equiv {\ln 4-1 \over 8\pi^2}N   + {\Delta v \over \sqrt{(\Delta y)^2+(\Delta v)^2}}{l_s\over y_{NS}^2} {1\over |h^2\mu|^2}.
\end{equation}
The first term in the parenthesis is dominant when $\mu \ll 1/l_s$. 

%%%%%%%%%%%%%%%%%%%%%%%%%%%%
\begin{figure}[htbp]
\begin{center}
 \includegraphics[width=.6\linewidth]{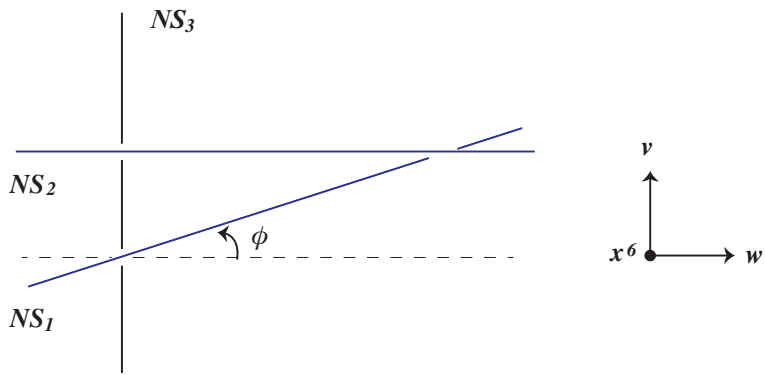}
\vspace{-.1cm}
\caption{\sl The $NS_1$ brane is rotated by the angle $\phi$ in $w-v$ plane.}
\label{brane2}
\end{center}
\end{figure}
%%%%%%%%%%%%%%%%%%%%%%%%%%%%

Now, let us engineer a SUSY preserving vacuum within the validity of this brane configuration without disturbing metastability. To do this, we slightly rotate the $NS_1$ brane by an angle $\phi$ in $(v,w)$ space. See figure \ref{brane2}. This rotation can be interpreted as adding the mass term of the moduli $X$ to the superpotential \cite{OoguriOokouchi,Franco,IAS,Giveon}, 
\begin{equation}
\Delta W={1\over 2} h^2 \mu_{\phi}  \tr X^2,\qquad \mu_{\phi}={\tan \phi \over 8\pi^2 g_s l_s}.\label{deform1}
\end{equation}
From the second expression in \eqref{deform1}, we see that the new parameter $\mu_{\phi}$ is described by geometric data. Adding this correction to the superpotential \eqref{magPotential}, we find that the SUSY preserving and breaking vacua are placed at 
\begin{equation}
X_{\rm SUSY}={\mu^2 \over h\mu_{\phi}},\qquad X_{\rm meta}\simeq {\mu_{\phi}\over h K_0}.
\end{equation}
From this, we see that the metastable vacuum is slightly shifted away from the origin. For later convenience, we define 
\begin{equation}
\Delta_8\equiv (X_{\rm SUSY}-X_{\rm meta})l_s^2,\qquad x_0^8\equiv X_{\rm meta}l_s^2.
\end{equation}

Finally, let us review the decay rate of the metastable vacuum. As in ISS model \cite{ISS}, we use the triangle approximation \cite{triangle} for
the evaluation since the depth of the potential near the metastable vacuum is
shallow but the distance from the metastable vacuum to the SUSY vacuum is
large. The transition can take place by creating a bubble in Minkowski space-time which corresponds to the domain wall $D4$ brane filling a subspace in the internal space. The tension of the bubble is 
\begin{equation}
T_{\rm DW}=T_{D4} S_2,\qquad S_2= \Delta_8 \sqrt{(\Delta_4)^2 +(\Delta y)^2}\simeq \Delta_8 \Delta_4,
\end{equation}
where $S_2$ is the area in $(x^4,x^6,x^8)$ filled by the $D4$ brane. For the sake of simplicity, we assume $\Delta y\ll \Delta_4$ below. The lifetime of the metastable vacuum is estimated by using the results in \cite{triangle}, 
\begin{eqnarray*}
\Gamma \sim \exp \left( -B_{\rm triangle}\right) ,\qquad  B_{\rm triangle}={2\pi^2\over 3} {(\Delta X)^4\over \Delta V  }.  \label{lifetime}
\end{eqnarray*}
In our setup, the distance between two vacua is given by $\Delta X \simeq X_{\rm SUSY}$. The vacuum energy of the metastable vacuum is 
\begin{equation}
\Delta V=V_{\rm peak}-V_{\rm meta}= N_{\rm peak} h^2\mu^4.
\end{equation}
Finding the peak of the potential is not a trivial task because in our approximation, this analysis is reliable only when
\begin{equation}
h\mu^2\gg h^2 \mu_{\phi} X.
\end{equation}
In other words, the behavior of the potential near the origin is
reliable. However, the peak of the potential places far away from the
origin, which makes the estimation slightly hard. Here, we roughly
assume the place of the peak is given by the geometric mean of two
vacua, $X_{\rm peak}={\mu \over h \sqrt{K_0}}$, and estimate the value
of $N_{\rm peak}$. Plugging the geometric mean back into the
potential, we get $V_{\rm peak} \sim 2V_{\rm meta} $, so we obtain
$N_{\rm peak}=1$. Hereafter, we simply assume that $N_{\rm peak}$ is
an ${\cal O}(1)$ constant.

%%%%%%%%%%%%%%%%%%%
\section{Fuzzy cosmic string (without gravity)}

%%%%%%%%%%%%%%%%%%%
\subsection{Dielectric tube-like brane nucleated from cosmic string \label{EnergyString}}

\begin{figure}[htbp]
\begin{center}
 \includegraphics[width=.5\linewidth]{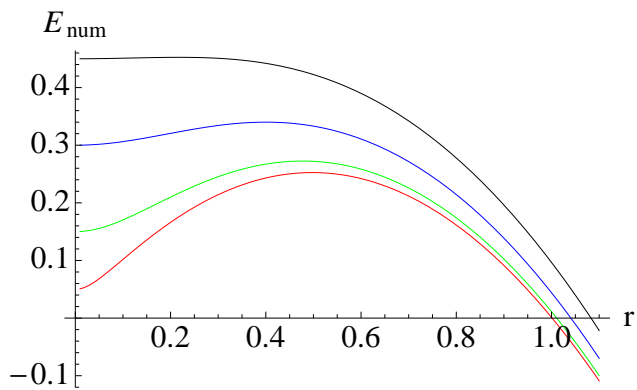}
  \vspace{0cm}
\caption{\sl $b=1/20,3/20,6/20,9/20$ for the red, green, blue and black lines}
\label{FIGpotmono}
\end{center}
\end{figure}

Now we are ready to discuss vacuum decay via a dielectric brane. In the
metastable vacuum there exists a string-like object firstly pointed out by
\cite{EHT}. A $D2$ brane connecting the $NS_1$ brane and the horizontal $D4$ branes
can be seen as a string in Minkowski space-time. See figure 1. So the
domain wall $D4$ brane and the $D2$
brane are stretching to the same directions in the internal space. Note that there is a small
displacement in $x^8$ direction that requires the extra cost to form the bound
state of $D2$/$D4$ branes \cite{Book}. However, since we mainly assume that $X_{\rm
SUSY}\gg X_{\rm meta}$ this energy cost can be negligible in our
argument below. By the $D2$ brane dissolving, the total energy of the system is given by 
\begin{eqnarray}
E_{\rm total}&=&\sqrt{\left(T_{D4}S_2 2\pi R L \right)^2  +\left(n_{D2}T_{D2} \Delta_8 L \right)^2  }-\Delta V \pi R^2 L \nonumber \\
&=& {4\pi T_{\rm DW}^2  L\over \Delta V} \left[ \sqrt{r^2+b^2} -r^2 \right] \nonumber \\
&\equiv & {4\pi T_{\rm DW}^2  L\over \Delta V}   E_{\rm num},
\end{eqnarray}
where we have defined,
\begin{equation}
r={ \Delta V \over 2T_{\rm DW}} R, \qquad b={n_{D2}T_{D2} \Delta_8 \Delta V\over 4\pi T_{\rm DW}^2}.
\end{equation}
Here, we have neglected the constant contribution from the homogeneous vev of
the metastable vacuum because it does not play any role for
dynamics. Also, we assumed that the $D2$/$D4$ brane forms a loop with
the size $L$. In figure \ref{FIGpotmono}, we show the energy
behavior. We find that there is no metastable dielectric brane with
finite size. The instability condition on the string-like object is 
\begin{equation}
b\ge {1\over 2}~,  \qquad {\rm (unstable)} .
\end{equation}

%%%%%%%%%%%%%%%%%%%%
%%%%%%%%%%%%%%%%%%%%

%%%%%%%%%%%%%%%%%%%%%%%%%%
\subsection{Decay rate of metastable string \label{DecayString}}

Now we have learnt the cosmic string can be either unstable or metastable depending on $b$.
For the parameter region $b\ge {1/2}$, even if the vacuum has enough
longevity, the vacuum decay via dielectric brane brings about the
instant phase transition. On the other hand, when the condition
$b<1/2$ is satisfied, the decay process is caused by an $O(2)$ symmetric instanton. In this subsection, we investigate this
decay process. Fortunately, we can proceed the study basically along the lines of \cite{HashII,hyakuII,PTKII}. 
We assume the following embedding function,
\begin{eqnarray}
&&X^0=t, \quad X^1=z, \quad X^2=R(t,z)\cos \theta ,\quad X^3=R(t,z)\sin \theta , \nonumber \\
&& X^4=x^4,\quad (0\le x^4 \le \Delta_4),\quad X^8=x^8,\quad (x_0^8\le x^8 \le x_0^8 +\Delta_8),
\end{eqnarray}
where $X^{5,6,7,9}$ are constant. Since the cosmic string is along $z$ and $x^4$ directions in the $D4$ brane coordinates, the dissolved $D2$-brane yields the magnetic field in the $(\theta, x^8)$ directions, Hence, the low energy effective action can be obtained by turning on $B\equiv 2\pi \alpha^{\prime}F_{\theta x^8}$. Exploiting the following expression,
\begin{equation}
\partial_{\alpha}X^{\mu}\partial_{\beta}X_{\mu}+2\pi \alpha^{\prime}F_{\alpha \beta}= \begin{pmatrix} % or matrix or bmatrix or Bmatrix or ... 
 -1+\dot{R}^2 & \dot{R}R^{\prime} &0& 0& 0 \\ 
 \dot{R}R^{\prime} & 1+R^{\prime 2} &0 & 0 & 0 \\
   0 & 0 & R^2  & 0 & B \\
   0 & 0 &0 & 1 & 0 \\
     0 & 0 &-B  & 0 & 1 
 \end{pmatrix},
\end{equation}
where $\alpha$ and $\beta$ are the indices of the world-sheet coordinates, the DBI action is given by 
\begin{equation}
S=-T_{D4} \int d^5 \xi \sqrt{(R^2+B^2)(1-\dot{R}^2+R^{\prime 2})} +\int
 dt\,  \pi R^2 L \Delta V .
\end{equation}
Below, for the sake of simplicity, we assume $R^{\prime}=0$. To
estimate the decay rate, let us write down the Euclidean action, 
\begin{equation}
S_E=\int d \tau \left[ 2\pi T_{\rm DW} L  \sqrt{(R^2+B^2)(1+\dot{R}^2)} - \pi R^2 L \Delta V \right],
\end{equation}
where $T_{\rm DW}=T_{D4}\Delta_4 \Delta_8$. It is useful to introduce the dimensionless parameters, 
\begin{equation}
s= { \Delta V \over 2 T_{\rm DW}}\tau  ,\qquad r= { \Delta V \over 2 T_{\rm DW}} R,\qquad b= { \Delta V \over 2T_{\rm DW}}B.
\end{equation}
Then, the Euclidean action becomes 
\begin{eqnarray}
S_E&=&2\pi  \left( {2 T_{\rm DW} \over \Delta V} \right)^2 T_{\rm DW}L \int ds \left[ \sqrt{(r^2+b^2)(1+\dot{r}^2)} -r^2\right] \nonumber \\
&=& {8\pi  L T_{\rm DW}^3 \over \Delta V^2} S_{\rm num}^{O(2)}.\label{E32}
\end{eqnarray}
The equation of motion can be described by the first order differential equation, 
\begin{equation}
\sqrt{{r^2+b^2\over 1+\dot{r}^2}}-r^2 =C,
\end{equation}
where $C$ is the integration constant. From this expression, we can discuss the velocity of a solution, $\dot{r}$, 
\begin{equation}
\dot{r}=\pm \sqrt{r^2+b^2-(C+r^2)^2\over (C+r^2)^2}.
\end{equation}
Since the initial condition is $\dot{r}=0$ at the core of the string $r=0$, the solution of the equation should satisfy the following factorization condition, 
\begin{equation}
r^2+b^2-(C+r^2)^2=r(r_{\rm max}-r)(r^2 +a_1 r+a_0).
\end{equation}
By solving this condition we find that $C=b$, $a_0=0$, $r_{\rm max}=a_1=\sqrt{1-2b}$. With this bounce solution, the velocity is written as follows, 
\begin{equation}
\dot{r}=\pm {r \sqrt{r^2_{\rm max}-r^2}  \over (b+r^2)}.
\end{equation}
Plugging this back into \eqref{E32}, we calculate the exponent of the decay rate, 
\begin{eqnarray}
 B_{O(2)}&=&2\cdot {8\pi  L T_{\rm DW}^3 \over \Delta V^2} \left(S^{O(2)}_{\rm num}(r_{\rm bounce})-S^{O(2)}_{\rm num}(0)\right) \nonumber \\
&=&{16\pi  L T_{\rm DW}^3 \over \Delta V^2} \left[  \int_{0}^{r_{\rm max}}dr {b+r^2 \over r\sqrt{r_{\rm max}^2-r^2}} \left( {r^2+b^2\over b+r^2} -r^2 \right) -b \int_{0}^{r_{\rm max}} dr {b+r^2 \over r \sqrt{r_{\rm max}^2 -r^2 }} \right] \nonumber \\ 
&=& {16\pi L T_{\rm DW}^3 \over \Delta V^2}\int_{0}^{r_{\rm max}} dr r \sqrt{r_{\rm max}^2 -r^2} \nonumber \\
&=& {16\pi L T_{\rm DW}^3 \over 3\Delta V^2} (1-2b)^{3\over 2} = B_{O(4)} l(1-2b)^{3\over 2},\label{B2/B4}
\end{eqnarray}
where $l\equiv 2\left({2\over 3} \right)^4 {\Delta V \over \pi T_{DW}}L$. Note that $B_{O(4)}$ is the bounce action of the thin-wall approximation \cite{Coleman}.  Since we are neglecting the curvature effect of the torus, the size $L$ should be much larger than the string scale $l_s$, $L\gg l_s$. Thus, within the validity of our approximation we find $l>1$. 

\begin{figure}[htbp]
\begin{center}
 \includegraphics[width=.5\linewidth]{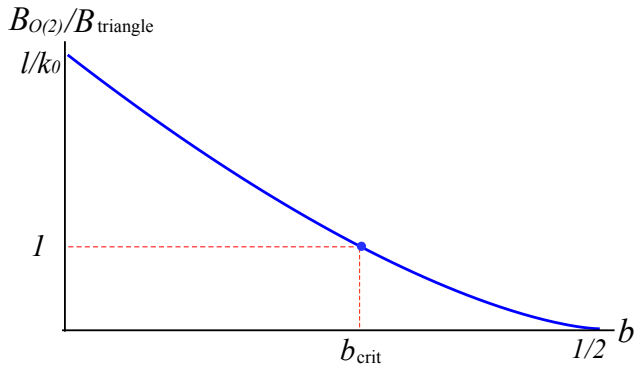}
  \vspace{0cm}
\caption{\sl The ratio of two bounce actions for the case of parameter choice, $l>k_0$.}
\label{FIGcritical}
\end{center}
\end{figure}

To estimate the decay rate, it is useful to rewrite $B_{\rm
triangle}$ in terms of $B_{O(4)}$. Recalling that the domain wall tension is written by $\Delta X$, we obtain
\begin{equation}
{B_{\rm triangle}\over B_{O(4)} }\equiv k_0= {4\pi^2\over 9N_{\rm peak}} \left( { \Delta V \over 3\pi (T_{D4} \Delta_4 l_s^2)^2 }\right)^2 <1.\label{Btri/B4}
\end{equation}
The last inequality comes from the fact that the depth of the
potential in the current setup is shallow and the triangle approximation is the dominant contribution. 
%With this relation, we see that dominant vacuum decay is different in
%two parameter regions. In the original setup, we assume that
%$B_{O(2)}$ is larger than $B_{\rm triangle}$, that means that the parameter $k_0$ is much smaller than unity.
From \eqref{B2/B4} and \eqref{Btri/B4}, we find that there is the critical point where two bounce actions become equal,  
\begin{equation}
k_0=l(1-2b_{\rm crit})^{3\over 2}.
\end{equation}
%From this critical point, $B_{\rm triangle}$ becomes larger than
%$B_{O(2)}$. 
As an illustration, we show a schematic behavior of the ratio of two bounce actions in figure \ref{FIGcritical}. In the region
$0\le b< b_{\rm crit}$, since $B_{O(2)}> B_{\rm triangle}$, the decay via
an $O(2)$ symmetric bubble is subdominant. On the other hand, In the
region $b_{\rm crit}< b<{1\over 2}$, we see that $B_{O(2)}<B_{\rm
triangle}$, so catalysis induced by the string plays an important role.

%%%%%%%%%%%%%%%%%%%%
%%%%%%%%%%%%%%%%%%%%

%%%%%%%%%%%%%%%%%%%%%%%%%%
\subsection{Adding fundamental string }

Finally, let us briefly study effects of colliding fundamental
strings to the fuzzy brane. First of all, we discuss the case when a fundamental string wraps in the
$\theta$ direction, namely wraps on the smaller cycle of the torus. In this case, doing the same way as before, we obtain the Lagrangian written in terms of the electric density flux $D$ as follows:
\begin{equation}
{S}= -\int dt \, \left[   2\pi T_{\rm DW}L\sqrt{(1+D^2)(R^2+B^2)(1-\dot{R}^2)} -\pi R^2 L\Delta V \right].
\end{equation}
By simply replacing the domain wall tension as $T_{\rm
DW}\to T_{\rm DW}\sqrt{1+D^2}$, we can apply the analysis done in the
subsection \ref{DecayString}. Since the tension becomes large, stability of the dielectric brane enhances. Physical meaning of this
effect is clear: wrapped fundamental strings try to shrink and generate forces making the radius of the dielectric brane small.

\begin{figure}[htbp]
\begin{center}
 \includegraphics[width=.5\linewidth]{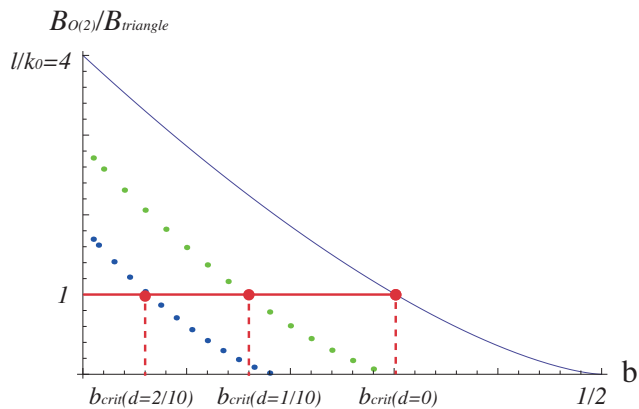}
  \vspace{0cm}
\caption{\sl The ratio of two bounce actions for $l/k_0=4$. Blue and green dots correspond to $d=2/10$ and $d=1/10$. When $b>b_{\rm crit}$, decays through $O(2)$ symmetric instantons are dominant. }
\label{FIGcritical2}
\end{center}
\end{figure}

The other interesting situation is that fundamental strings are
dissolving in the domain wall brane along the $x^8$ direction. In this
case, by calculating the DBI action explicitly, we find again that the
action, for $R^{\prime}=0$, becomes exactly the same as the one
studied in \cite{KO}. The electric flux is along the $x^8$ direction
while the magnetic flux is in $(\theta, x^8)$ directions, the total action of the tube-like brane becomes 
\begin{equation}
S=-\int d t\left[ 2\pi L T_{\rm DW}  \sqrt{R^2(1-\dot{R}^2-E_8^2)+B^2(1-\dot{R}^2)} - \pi  R^2 L \Delta V\right].
\end{equation}
Using the results shown in \cite{KO}, we find that the critical value depends on the strength of the electric flux. The result is plotted in figure
\ref{FIGcritical2}. Finally, when the fundamental string dissolves along the $z$-direction, we can easily obtain the decay rate by replacing $B^2\to B^2+D^2$.

%%%%%%%%%%%%%%%%%%%%
%%%%%%%%%%%%%%%%%%%%

%%%%%%%%%%%%%%%%%%%%%%%%
\section{Gravitational corrections}

In this section, we discuss gravitational interactions in ten-dimension between the
branes
and the fuzzy brane. As studied in \cite{Kutasov}, the interactions
generate a non-negligible effective potential for light fields and
modify the landscape of the potential. Here, we study dynamics of the fuzzy brane under such interactions. This is the first step to get insights into gravitational corrections for inhomogeneous vacuum decay in string landscape.

%%%%%%%%%%%%%%%%%
\subsection{Stability of cosmic string corrected by gravity}

At first, we review stability of the false vacuum under such gravitational interactions. The $NS_1$, $NS_2$ branes and the tilted $D4$ branes are mutually BPS, so there is no force
between them. On the other hand, since the $NS_3$ brane and the tilted $D4$
branes are mutually non-BPS, they interact non-trivially, which leads to the most important correction in the
following argument. When the distance of the $D4$-brane from the $NS_3$ brane is quite large, this correction can be treated as
 the classical background five-brane solution \cite{Kutasov, strominger}. We can take account of the
gravitational correction to the DBI action by analyzing the motion of the $D4$
brane in the presence of this background. Also, the tilted $D4$ branes and the horizontal $D4$ branes are mutually non-BPS, so
there is a force between them which has been already incorporated as loop corrections of open strings between them.

The classical
solution of the $NS$-five brane discussed in \cite{strominger} is
\begin{eqnarray}
&&d s^2=d x_{\mu} d x^{\mu} +d v d \bar{v} +H(r_{\rm NS}) [dy^2 +(dx^7)^2+d w d \bar{w}],\nonumber \\
 && e^{2(\varphi-\varphi_0)}=H(x^n) \\
 && H_{lmn}=-\epsilon_{lmn p}\partial_{q} \varphi.\label{NSsolution}
\end{eqnarray}
$H_{lmn}$ is the strength of the NS $B$ field, $dB_2$. The indices run $\mu,\nu=0,1,2,3$ and
$l,m,n,p=6,7,8,9$. The exponential of dilaton, $e^{\varphi_0}$, is interpreted as the string
coupling constant $g_s$. The harmonic function $H$ is given by
\begin{equation}
H(r_{\rm NS}) =1+{l_s^2 \over r^2_{\rm NS}},\qquad {\rm where} \quad r^2_{\rm NS}=(y-y_{NS})^2 +(x^7)^2+|w|^2.
\end{equation}
As mentioned above, since the dominant gravitational correction comes
from the $NS_3$ brane, so we apply this solution only to the $NS_3$ brane and
study dynamics of the $n$ tilted $D4$ branes and the fuzzy $D4$
brane. By gravitational corrections, the minimum of the SUSY breaking vacuum shifts slightly. To see this, write down the DBI action of the
tilted
$D4$ brane in this background. The potential for the $D4$ brane becomes \cite{Kutasov}
\begin{equation}
V={T_4 \Delta y \over g_s}\left( \sqrt{1+{(2\pi)^4(l_sh\mu)^2 \over H(r_{\rm NS})}}-1\right) .
\end{equation}
This depends on $w$, and we find that there is a force pulling the $D4$
brane back to the origin. When the $NS_3$ brane is far enough from the other $NS$
branes, $y_{NS}\to -\infty$, the energy of the metastable
vacuum is given as the minimum of this potential plus the correction
mentioned in the previous section, $V_{\rm correction}$. We do not
need the actual
value of this in the following discussion, then we formally write this minimum and
the energy at the minimum as
\begin{equation}
w_{\rm meta}=x_{\rm meta}, \qquad \Delta V={T_4 \Delta y \over g_s}\left( \sqrt{1+{(2\pi)^4(l_sh\mu)^2 \over H(x_{\rm meta})}}-1\right) .
\end{equation}
Here we have fixed the energy of the SUSY vacuum to be 0.

\begin{figure}[htbp]
\begin{center}
 \includegraphics[width=.42\linewidth]{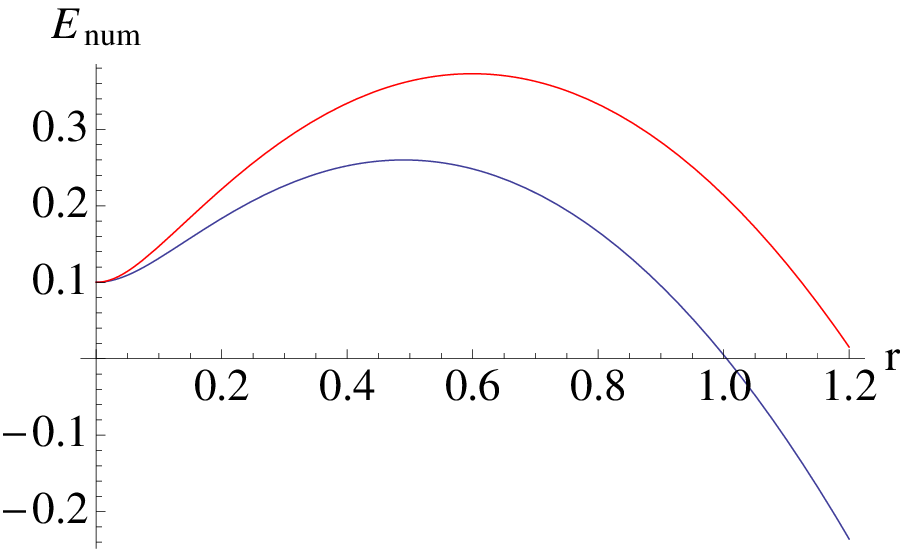}\hspace{1cm}
  \includegraphics[width=.42\linewidth]{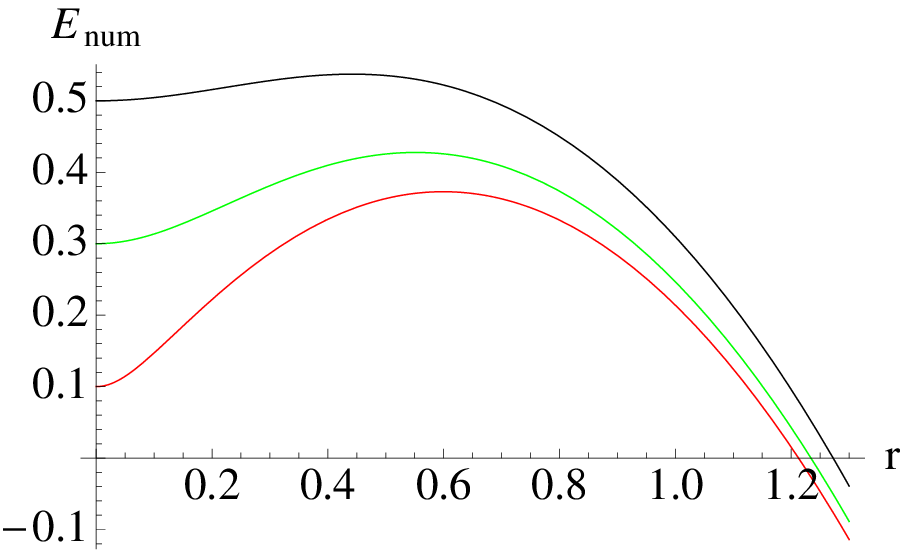} 
  \vspace{0cm}
\caption{\sl In the left panel. Red for $\Delta Y=5$, $L_s=5$, $b=1/10$, $\epsilon=1/10$. Blue corresponds to the case with $L_s=0$. In the right panel, $\Delta Y=5$, $L_s=5$, $\epsilon=1/10$ the red, green and black correspond to $b=1/10$, $3/10$ and $5/10$.}
\label{AAA}
\end{center}
\end{figure}

Now we have reviewed gravitational corrections to the metastable
vacuum, we start to think gravitational corrections to the vacuum
decay via the cosmic string. As in the previous section, suppose there
is a string-like object corresponding to the $D2$ brane stretched
between the $NS_1$ brane and the horizontal $D4$ branes, and assume that a tube-like domain wall connecting two vacua is induced by the string. The string on the top of the $D4$ brane is energetically unstable and dissolves into the $D4$ brane \cite{Book}. We investigate dynamics of the $D2$/$D4$ bound state. We take the embedding function of the $D4$ brane the same as in the previous section. In the background \eqref{NSsolution}, the DBI action of the $D4$ brane becomes
\begin{equation}
\sqrt{-\det (\partial_{\alpha} X^{\mu} \partial_{\beta}X_{\mu}+2\pi
\alpha^{\prime}F_{\alpha \beta} )}=\sqrt{H(r_{\rm NS})R^2 +B^2},
\end{equation}
where we have assumed that the radius $R$ is constant along the string and time. We find the potential energy to be
\begin{equation}
E=\left( T_{D4}\Delta_4 \Delta_8\right) {1\over \Delta_8} \int dx^8\, 
 2\pi L \sqrt{H(r_{\rm NS})R^2 +B^2}-\pi R^2 L \Delta V .
\end{equation}
It is useful to introduce dimensionless variables,
\begin{equation}
x^8\equiv x_{\rm meta}\cdot q,\qquad \epsilon\equiv {x_{\rm meta}\over \Delta_8}, \qquad L_s\equiv {l_s \over  x_{\rm meta}},\qquad  \Delta Y\equiv {\Delta y\over x_{\rm meta}}.
\end{equation}
For the sake of simplicity, we assume that $x^7$ and $\Delta y$ dependences are negligible. So, we have $r^2_{\rm NS}=(\Delta y_1)^2 +(x^8)^2$. 
By plugging these expressions, the energy function is written as follows:
\begin{equation}
E=2\pi T_{\rm DW} \left({2T_{\rm DW}\over \Delta V} \right) L  \left[   \epsilon \int_{1}^{1+1/\epsilon} dq \sqrt{ \left(1+{L_s^2\over \Delta Y^2 +q^2 } \right)  r^2 +b^2}-r^2\right].
\end{equation}
We have used the fact $x_{\rm SUSY}=x_{\rm meta}+\Delta_8$. As an illustration, we plot
this potential in figure \ref{AAA}. Obviously, the potential barrier
becomes higher by gravitational corrections, thus the metastable state
is stabilized. This can be understood that a force generated by the black $NS$-five brane pulling back the domain
wall $D4$ brane impedes the
decay\footnote{Note that when we add corrections arising
from $H$, as we discussed earlier, $\Delta V$ becomes slightly lower than the one in the previous
section. Taking account of this, the false vacuum is more stable
than without the corrections.}.

%%%%%%%%%%%%%%%%%%%%
%%%%%%%%%%%%%%%%%%%%

From this potential, we can read off the critical strength of the magnetic field that makes
the fuzzy brane unstable. Differentiating $E$ with respect to $r$, we find that
the local minimum exists at $r=0$, and the maximum at the radius satisfying the following condition,
\begin{equation}
\epsilon \int_{1}^{1+1/\epsilon} dq \left[ {H \over \sqrt{H r^2 +b^2}}-2\right]=0 .
\end{equation}
In our assumption, the harmonic function is given by 
\begin{equation}
H=1+{L_s^2 \over \Delta Y^2 +q^2} .
\end{equation}
At the critical strength of the magnetic field these two points collide. This
condition is written as
\begin{equation}
2b_{\rm unst}=\epsilon \int_{1}^{1+1/\epsilon} H dq .
\end{equation}
This integral is analytically solved and we find the critical value is described by
\begin{eqnarray}
b_{\rm unst}&=&{1\over 2}+{\epsilon L_s^2 \over 2 \Delta Y} \left[ \tan^{-1}\left({1\over \Delta Y}+{1\over \epsilon \Delta Y} \right)-\tan^{-1} \left({1\over \Delta Y} \right) \right] \nonumber \\
&=& {1\over 2}+{\epsilon L_s^2 \over 2 \Delta Y} \tan^{-1}
 \left({\Delta Y \over \epsilon \Delta Y^2 +\epsilon+1} \right) .
\end{eqnarray}
In the limit $L_s\to 0$, this result reproduces the critical value shown in previous section. From the positivity of the second term, we can read off the tendency that the gravitational correction stabilizes the potential and that the critical value of the magnetic field becomes larger.

%%%%%%%%%%%%%
\subsection{How decay rate changes with gravity}

\begin{figure}[htbp]
 \begin{center}
 \includegraphics[width=.5\linewidth]{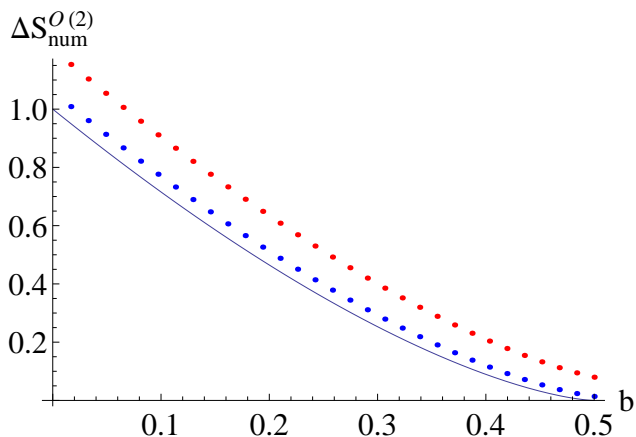}
  \vspace{0cm}
\caption{\sl Numerical plots of $\Delta S_{\rm num}^{O(2)}$ in \eqref{Added} for the parameters  $\epsilon=1/100$, $\Delta Y=9$ and $L_s=5$ (blue dots) and $\epsilon=1/100$, $\Delta Y=9$ and $L_s=9$ (red dots) . The blue line corresponds to $\epsilon=1/100$, $\Delta Y=9$ and $L_s=0$.}
\label{FIGgravitation}
 \end{center}
\end{figure}

We have studied how gravitational corrections change the energy of the false vacuum
and the critical value of the magnetic field. Also, we have studied existence of cosmic strings makes change
of the decay rate in the previous section (see figure \ref{FIGcritical}). Now we are ready to investigate how gravity affects on
this phenomenon by evaluating the decay rate in a similar way. We start from writing down the
Euclidean Lagrangian that respects the gravitational correction,
\begin{equation}
S_E= \int d \tau \left[ 2\pi T_{\rm DW}L {1\over \Delta_8}\int d x^8
		  \sqrt{(1+\dot{R}^2)(HR^2+B^2)}- \pi R^2 L \Delta V
		 \right] .
\end{equation}
Again, introducing dimensionless variables
\begin{equation}
\tilde{b}={\Delta V\over 2HT_{\rm DW}}B, \qquad r={ \Delta V \over 2\sqrt{H} T_{\rm DW} }R,\qquad s={ \Delta V \over 2\sqrt{H} T_{\rm DW} } \tau,
\end{equation}
we obtain
\begin{equation}
S_E={8\pi T_{\rm DW}^3L \over \Delta V^2 } \int ds {1\over \Delta_8}\int_{x_{\rm
 meta}}^{x_{\rm meta}+\Delta_8} d x^8 \,  H^{3/2}  \left[
						       \sqrt{(1+\dot{r}^2)(r^2+\tilde{b}^2)}-
						       r^2 \right] .
\end{equation}
We wrote the second term in the form of integral with respect to $x^8$ as well as
the first term. With the integration over $x^8$, it seems difficult to
calculate the bounce action analytically. However, note that $x^8$ is time independent. Therefore
we can fix $x^8$ once and calculate the bounce action in the same way as the previous section, then we integrate the analytic result over $x^8$. In this way, we obtain the bounce action
\begin{eqnarray}
B_{O(2)}^{\rm grav}&=& {B_{O(4)}l\over \Delta_8}\int dx^8 H^{3/2}
 (1-2\tilde{b})^{3\over 2}= {B_{O(4)}l\over \Delta_8}\int dx^8 H^{3/2}
 \left(1-2{b\over H}\right)^{3\over 2}  \nonumber \\
 &=&{B_{O(4)}l\over \Delta_8}\Delta S^{O(2)}_{\rm num} . 
\end{eqnarray}
It is convenient to evaluate the integral, $\Delta S_{\rm num}^{O(2)}$ numerically,
\begin{equation}
\Delta S^{O(2)}_{\rm num}=\int_{1}^{1+1/\epsilon} dq  \left(H-2b\right)^{3\over 2},\quad {\rm where} \quad H= \left(1+{L_s^2\over \Delta Y^2+q^2} \right) .\label{Added}
\end{equation}
We plot this in figure \ref{FIGgravitation}.
We may say the gravitational corrections stabilize the bubble.

Note that this
calculation is able to do only when the bounce action can be
evaluated analytically. If not, we have to numerically evaluate the
bounce action with fixed $x^8$. However, $\tilde{b}$ depends on $x^8$, so we cannot integrate over $x^8$ after
that. In the next subsection, we will meet one of such examples, which requires further assumption to proceed calculations. 
%On the other hand, in the case that
%a fundamental string winding on the torus, probably we are able to evaluate the bounce action by
%just correcting the variables by multiplying the factor in the
%result without the gravitational correction. This is also a curious incident.

%%%%%%%%%%%%%%%%%%%%
%%%%%%%%%%%%%%%%%%%%

%%%%%%%%%%%%%%%%
\subsection{Adding fundamental string again}

\begin{figure}[htbp]
\begin{center}
 \includegraphics[width=.5\linewidth]{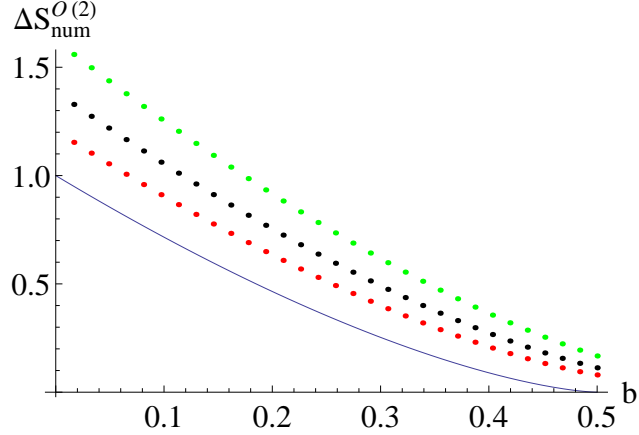}
  \vspace{0cm}
\caption{\sl We take $\Delta Y=9$, $L_s=9$ and $\epsilon=1/100$. The red, black and green dots correspond to $D=0,1/3$ and $1/2$. }
\label{FIGgravitation2}
\end{center}
\end{figure}

For completeness, let us again discuss the case with fundamental strings winding on the torus. At first, we deal with fundamental strings winding in $\theta$ direction. We take the same coordinates as before, and the fundamental strings
dissolve and induce the electric field toward $\theta$ direction on the domain wall $D4$ brane. Then, the DBI action becomes
\begin{equation}
S=\int dt \left[ -2\pi LT_{DW} {1\over \Delta_8}\int dx^8
	   \sqrt{(H(r)R^2+B^2)(-1+\dot{R}^2)-E^2 H(r) } +\pi R^2 L
	   \Delta V   \right] .
\end{equation}
We make Legendre transformation and rewrite with
the electric density flux $D$. Euclideanizing this, we get 
\begin{equation}
S_E=\int dt \left[ 2\pi LT_{DW} {1\over \Delta_8}\int dx^8
	     \sqrt{\frac{H+D^2}{H}} \sqrt{(H(r)R^2+B^2)(1+\dot{R}^2)} -\pi
	     R^2L\Delta V \right] .
\end{equation}
Using the following dimensionless variables
\begin{eqnarray}
&&r={ \Delta V \over 2T_{\rm DW}\sqrt{H+D^2}} R,\qquad s={ \Delta V \over 2T_{\rm DW}\sqrt{H+D^2}} \tau,\qquad \tilde{b}={\Delta V\over 2T_{\rm DW}\sqrt{H}\sqrt{H+D^2} }B,\nonumber \\
 &&l\equiv
 \left({2\over 3}\right)^4{\Delta VL\over \pi T_{DW}} ,
\end{eqnarray}
the action can be written as
\begin{equation}
S_E={l\over 2} B_{O(4)} \int ds \left[ {1\over \Delta_8}\int d x^8
				 \left(H+D^2
				 \right)^{3/2}\left(\sqrt{(1+\dot{r}^2)(r^2+\tilde{b}^2)}
				 -r^2 \right) \right] .
\end{equation}
As in the previous section, we can easily find the critical value of the magnetic field beyond which the dielectric brane becomes unstable, 
\begin{equation}
b_{\rm unst}={\int dx^8 \sqrt{H}(H+D^2)^2\over \int dx^8 (H+D^2)^{3/2}}.
\end{equation}
Again, $x^8$ is independent of the time. We may find the bounce action with fixed $x^8$ first and then integrate over it after that,
\begin{eqnarray}
B_{O(2)}^{\rm grav}&=&{1\over \Delta_8}\int d x^8 l \left(H+D^2 \right)^{3/2}B_{O(4)}\left(1-2\tilde{b}  \right)^{3\over 2}  \nonumber \\
&=&{1\over \Delta_8}\int d x^8 l \left(H+D^2
				 \right)^{3/2}B_{O(4)}\left(1-{2{b}\over
				 \sqrt{H(H+D^2)}}  \right)^{3\over 2}
				 \nonumber \\
 &=&{l\over \Delta_8}B_{O(4)}\Delta S^{O(2)}_{\rm num} .
\end{eqnarray}
It is difficult to evaluate this analytically. So we plot the
numerical results in figure \ref{FIGgravitation2}. Again, we see that
the large electric flux stabilizes the bubble.

%%%%%%%%%%%%%%%%%%%%
%%%%%%%%%%%%%%%%%%%%

\begin{figure}[h]
\begin{center}
 \includegraphics[width=.45\linewidth]{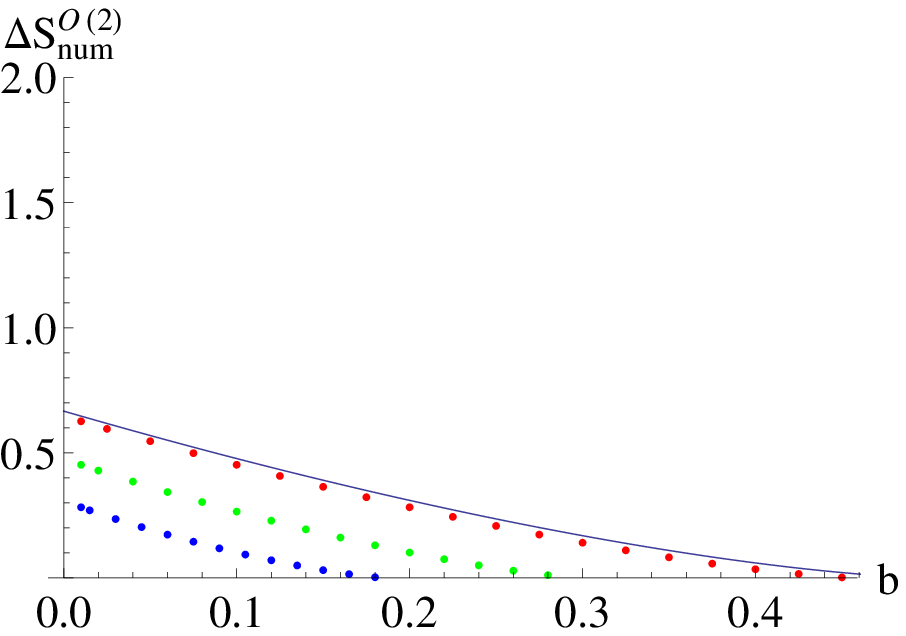}\hspace{0.7cm}
  \includegraphics[width=.45\linewidth]{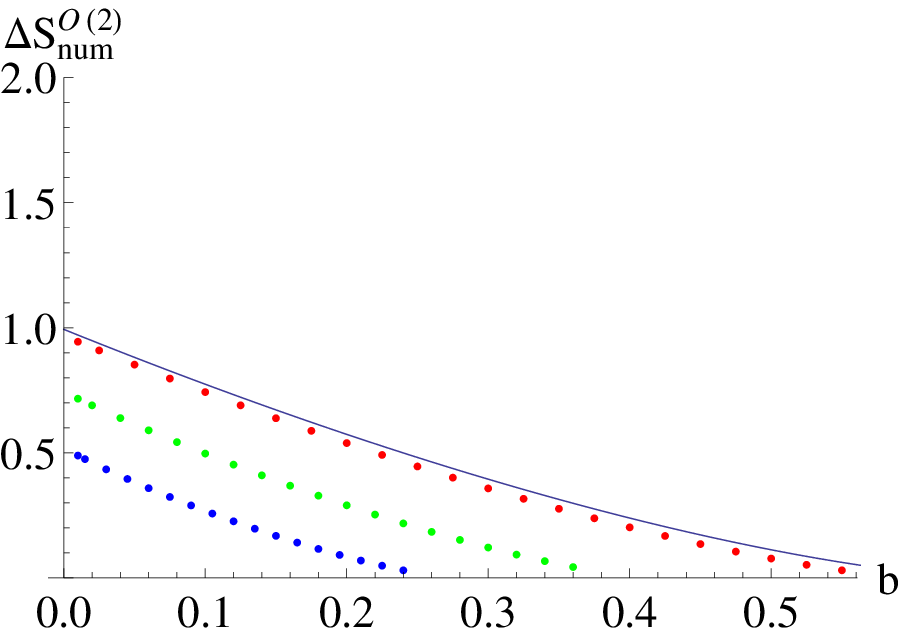}\hspace{0.7cm}
 \includegraphics[width=.45\linewidth]{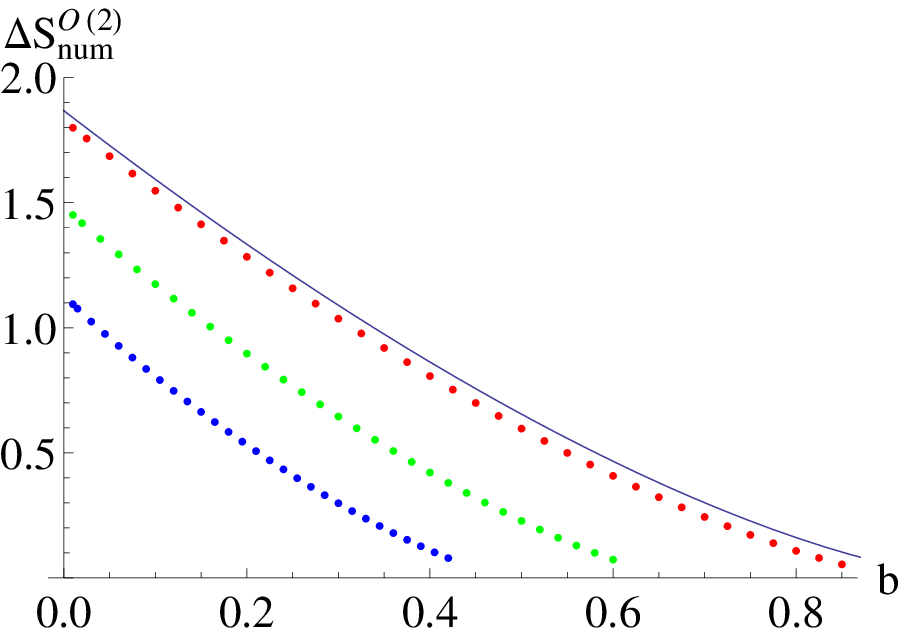}
  \vspace{0cm}
\caption{\sl In each figure, the blue line corresponds to $d=0$. The red, green, and blue dots
 correspond to $d=1/100$, $d=1/10$, $d=2/10$ respectively. In the upper left
 panel, we take $\Delta Y=0, L_s=0$ (no gravity). In the upper right panel, we take
 $\Delta Y=9, L_s=5$. In the lower panel, we take $\Delta Y=9, L_s=9$.}
\label{gratorus}
\end{center}
\end{figure}

Finally, we briefly discuss effects adding the fundamental strings in $x^8$
direction. As we have done before, we find the Euclidean action after
Legendre transformation to be
\begin{equation}
S_E=\int dt \left[ 2\pi LT_{DW} {1\over \Delta_8}\int dx^8  {1\over
	     R}\sqrt{(HR^2+B^2)(1+\dot{R}^2)(R^2+D^2)} -\pi R^2 L
	     \Delta V   \right] .
\end{equation}
Unfortunately in contrast to the previous cases, it is difficult to evaluate the bounce action even with fixed
$x^8$. Thus, we have to require one more approximation and try to see the tendency of
the gravitational correction to the bounce action. Suppose that $H$ is almost
independent of $x^8$. This assumption limits the effective parameter
region to be rather small. The harmonic function becomes
\begin{equation}
H_0\equiv 1+{L_s^2 \over \Delta Y^2+q_{\rm meta}^2}.
\end{equation}
By integrating over $x^8$, we obtain the simple expression
\begin{equation}
S_E=\int dt \left[ 2\pi LT_{DW}  {1\over R}\sqrt{(H_0R^2+B^2)(1+\dot{R}^2)(R^2+D^2)} -\pi R^2 L \Delta V   \right].
\end{equation}
We use dimensionless variables
\begin{eqnarray}
s={ {\Delta V}\over 2\sqrt{H_0}T_{DW} } \tau,\quad r={ {\Delta V}\over 2\sqrt{H_0}T_{DW} } R,\quad \tilde{b}={ {\Delta V}\over 2H_0T_{DW} } B,\quad 
 \tilde{d}={ {\Delta V}\over 2\sqrt{H_0}T_{DW} } D.
\end{eqnarray}
Then,
\begin{eqnarray}
  S_E&=&\int ds {8\pi L T_{DW}^3 \over {(\Delta V)^2}}H_0^{3/2}
  \left[{1 \over r} \sqrt{(r^2+\tilde{b}^2)(1+\dot{r}^2)(r^2+\tilde{d}^2)}-r^2
 \right]  \nonumber \\
 &=&\int ds {8\pi L T_{DW}^3 \over {(\Delta V)^2}}H_0^{3/2}
  \left[{1 \over r} \sqrt{(r^2+{b^2 \over
   {H_0^2}})(1+\dot{r}^2)(r^2+{d^2 \over {H_0}})}-r^2
 \right] \nonumber \\
     &=&{8\pi L T_{DW}^3 \over {(\Delta V)^2}} S_{\rm num}^{O(2)}
      \nonumber \\
 &=&3l B_{O(4)} S_{\rm num}^{O(2)},
\end{eqnarray}
where we have defined $ l\equiv
 \left({2/ 3}\right)^4{\Delta VL/ \pi T_{DW}}$. By defining the rescaled  variables, $b/H_0$ and $d/\sqrt{H_0}$, the action becomes the same as the one in \cite{KO} up to overall factor $H^{3/2}_0$. Hence, we can easily evaluate the bounce action using the results in \cite{KO}.
% 
% In this case, the gravitational correction is just an over all scaling by $H_0^{3/2}$. Thus we calculate the decay rate by quite similar way in \cite{KO} in which
%the decay rate without the gravitational correction is evaluated.
For the numerical estimation, it is useful to rewrite the bounce action formally as follows:
\begin{eqnarray}
 B_{O(2)}^{\rm grav}&=&{16\pi L T_{DW}^3 \over {(\Delta V)^2}} \Delta
  S_{\rm num}^{O(2)}  \nonumber \\
 &=&6l B_{O(4)} \Delta
  S_{\rm num}^{O(2)} .
\end{eqnarray}
We plot $\Delta S_{\rm num}^{O(2)}$ as a function of $b$ in figure \ref{gratorus}.
Again we have found the tendency that the gravitational correction
stabilizes the bubble, even with a string winding toward $x^8$ direction.
%%%%%%%%%%%%%%%%%%%%%%%%%%%%%%%
%%%%%%%%%%%%%%%%%%%%%%%
\section{Conclusions and discussions}

In this paper, we have studied transitions between false and true vacua. These vacua were engineered
by brane configurations in Type IIA string theory. We discussed the
inhomogeneous vacuum decay via a $O(2)$ symmetric bubble induced by a
cosmic string in the false vacuum. In the decay process, the bubble
and the string formed a bound state which corresponds to the
dielectric brane in the context of string theory. We found that the
lifetime of the false vacuum becomes shorter in a wide range of
parameter space. However, remarkably, when the induced magnetic field
on the dielectric brane is smaller than the critical value, this catalysis does not work. This is in contrast to the results in the first paper \cite{KO}. We also discussed the leading order gravitational corrections in ten-dimension by treating the $NS_3$ brane as the background metric of the extremal black brane solution. This correction induced
the force (basically) pulling the domain wall back to the
origin in the internal space. And this force made the potential barrier higher and therefore
the decay rate became smaller.

%In evaluating this decay
%rate, there are two parameter regions. One is that in which the string
%winding number is small and the vacuum is not always stabilized by existence of
% the cosmic string. The other is the case with large string winding number in
% which the strong magnetic field exists on the domain wall and the decay
% via soliton occurs as well as in \cite{KO}. Also, we discussed the case a
% fundamental string winding on the bubble and concluded that this
% winding changes the stability of the bubble. If the fundamental
% string winds along the shorter circle of the torus, because of
% tension of the string, the tube-like bubble is stabilized. On the
% other hand, if the string winds on the longer circle, this causes the
% false vacuum to decay more quickly.

%
%This
%tendency means that if string coupling constant is large, it is not
%obvious that this decay process is the dominant one. In this paper, we
%discussed gravitational corrections in a special setup. For more
%general discussion, we have to do the same analysis in different models
%and check generality.

%As an advanced discussion, we can think the NS-five brane as a non-extremal
%black hole, and this is equivalent to think the finite temperature
%system.

To proceed our studies further, it would be important to incorporate finite temperature effects. In the present brane setup, the finite temperature effects can be treated by using the non-extremal black brane solutions \cite{KutasovFinite}. However, the analysis would become quite involved by the following two reasons. One is that since all $NS$-five branes are replaced by the black brane
solutions, the original metastable configuration itself is distorted by the temperature effects. Therefore existence of the metastable state itself is not obvious in
some parameter region. The other reason is that in a finite temperature
system, the states are given by the Boltzmann distribution, so the initial
point of the decay is not always the minimum of the potential. Thus we have to
evaluate the thermally assisted decay rate and this analysis makes
evaluation of the bounce action complicated. Also, to show generosity
of inhomogenious decay in string landscape, it would be interesting to
extend our studies of gravitational effects to geometrically induced
metastable vacua \cite{Geometry,SeibergWitten}. However, these studies
are beyond the scope of this paper, so we would like to leave them future works.

%%%%%%%%%%%%%%%%%%%%%
\section*{Acknowledgement}

The authors would like to thank Yuichiro Nakai for comments and discussions. The authors are grateful to Harvard University for their kind hospitality where this work was at the final stage. This work is supported by Grant-in-Aid for Scientific Research from the Ministry of Education, Culture, Sports, Science and Technology, Japan (No. 25800144 and No. 25105011).

%%%%%%%%%%%%%%%%%%%%%%%%%%%%%%%%%%%%%%%%%%%%%%
%
% Bibliography
%
%%%%%%%%%%%%%%%%%%%%%%%%%%%%%%%%%%%%%%%%%%%%%%

\end{document}